# Technical Report

## Title: SDP-based State Estimation of Multi-phase Active Distribution Networks using micro-PMUs

**Authors:**

*Vahid Rasouli Disfani,*

*Mohammad Chehreghani Bozchalui*

*Ratnesh Sharma*

**Abstract-** Distribution system state estimation (DSSE) is an essential tool for operation of distribution networks, the results of which enables the operator to have a thorough observation of the system. Thus, most distribution management systems (DMS) include a single-phase state estimator. However, single-phase state estimation is not able to address the unbalanced nature of a distribution system which led to the concept of three-phase DSE in the literature and industry. Besides, transition of distribution systems from passive to active ones that include Distributed Energy Resources (DER) necessitates more comprehensive state estimation techniques. Due to non-convexity of the SE problem, heuristic and Newton's methods do not guarantee the global solution. In contrast, SDP based SE is more promising to guarantee the globally optimal solution since it represents and solves the problem in a convex format. However, the observability of the power system is highly vulnerable to the set of measurements while employing the SDP-based SE, which is addressed in this report. An algorithm is proposed to generate additional measurements using the measurement data already gathered. The SDP-based SE is very sensitive to the level of noise in large power networks. Also, bad data detection algorithms proposed for Newton's methods do not work for the SDP-based SE method due to larger number of state variables in SDP representation of power network. In this report, an algorithm is proposed to generate additional measurements using the measurement data already gathered in order to solve the observability issue. A network separation algorithm is developed to solve the entire problem for smaller sub-networks which include $\mu$PMUs to mitigate the adverse effects of noise for huge networks. An algorithm based on redundancy test is developed for bad data detection. The algorithms are tested on single phase and multiphase test systems. The algorithms are applied EPRI Circuit 5 (2998-bus) test feeder to demonstrate the flexibility of the algorithms developed.

# Introduction

## Distribution System State Estimation

Electrical distribution system is a part of power systems which is directly connected to the end users. It is crucial to have a real-time monitoring over all nodes of the distribution system to guarantee reliable and high-quality services to the customers. Unfortunately, it is not feasible to connect sensors on all nodes of the distribution system due to their costs. However, Distribution System State Estimation (DSSE) is employed to provide the Distribution Management Systems (DMS) with thorough information about the state of the distribution systems using a limited number of measurement gathered from the network [1].

The concept of State Estimation (SE) was first developed for transmission systems known as Power System State Estimation (PSSE) [2]–[4]. The idea is to define the most possible states of the power systems (voltage magnitudes and angles of all nodes) which would induce the actual measurements obtained by the sensors.

In PSSE, transmission systems are typically modeled using positive sequence equivalent of the system for simpler and faster analyses. This simplification is acceptable for transmission grid because large generators and transmission lines are inherently symmetric and operated in a balanced condition. On the other hand, distribution systems are unbalanced three-phase networks with single-phase and two-phase feeders, loads, and distributed generation resources. Also, distribution networks have small ratios of X/R, large numbers of load points, un-transposed lines with phase impedances, and less redundancy of measurements in distribution system [5]. Therefore, the DSSE imposes a high-dimensional mathematical problem and is generally more difficult to be modeled than PSSE.

## Background

Both PSSE [1], [6]–[18] and DSSE [19]–[28] have been reported in enormous volume of research works in the literature. SE results are used for several applications by power system management systems such as fault detection [29] and optimal control of DERs [30]. The most common estimator for SE is Weighted Least Square (WLS) [1], [9]. A comprehensive comparison of various estimators for DSSE is provided in [25].

One of the major challenges for SE problem is the inherent non-convexity of the SE problem due to the nonlinear model of power systems [16]–[18]. It makes the globally optimal solution difficult to be sought. The most widely used algorithm to solve SE optimization problem is Newton's method, but it is subject to be stuck in local optima instead of converging to globally optimal solution. In fact, Newton's method is very sensitive to the initial guess which has a key role to define the final solution. Heuristic methods such as particle swarm optimization [31] are the other options to solve the SE problem. Heuristic methods are not guaranteed to converge to the global optimum in spite of their remarkable speed. Semidefinite Programming (SDP) based approaches are proposed for PSSE in [16]–[18]. These methods are based on a SDP representation of the power system equations [32]. By relaxing the rank constraint in the SDP problem, the SE is converted to a convex optimization problem, which can be solved effectively to find the global optimum. If the rank constraint in the SDP-based SE solutions is not satisfied, these results can be used as initial guesses for Newton's based PSSE method to find the optimal solution [17].

DSSE is more challenging than PSSE due to higher untransposed lines with phase imbalances, small X/R ratios, larger numbers of nodes and load points, and less redundancy from Kirchhoff's law [5], [33]. Distribution systems present a high dimensional mathematical problem while offering few physical measurements to be fed into DSSE algorithm. With the advent of smart grids, distribution systems are equipped with several types of Distributed Energy Resources (DERs) such as renewable energy resources, energy storage systems, demand response, and distributed generation. These resources may be connected to one, two, or three phases and make the distribution system more unbalanced. To handle the inherent imbalance of the distribution systems and DERs, development of a multiphase DSSE technique is vital. To model the DSSE problem, several methods have been reported in the literature. Branch-Current-Based DSSE algorithm [21], [22], [24] considers the current phasors of the branches as the state variables in the estimation problem. Some methods are proposed to solve PSSE faster by reducing the size of problem via network reduction [28], [33], [34]. Linear formulation of SE [26], which deals with real and imaginary parts of voltage phasors as states instead of their magnitudes and angles, is defined based on a linearized representation of the power system equations. DSSE for distribution networks including DERs have also been investigated in several articles [24], [26], [28], [33], [35].

Several methods have been proposed in the literature to incorporate the measurements gathered by Phasor Measurement Units (PMUs) in the SE problem [10], [12], [35]–[37]. PMUs are measurements units which provide synchronized phasor measurements of voltage and current signals. Although installation of PMU does not seem an economic solution for distribution system monitoring at this time, there are some ongoing research projects to introduce lower-cost PMUs called $\mu$PMUs to increase the observability of distribution systems [5].

The SE problem is always concerned of observability of the system investigated which is highly influenced by the configuration of the system and the measurement set [1], [38], [39]. Observability of distribution systems is highly affected by low redundancy of real measurements [5], [33]. Therefore, utilizing the historic data of distribution system, especially load profile, to generate pseudo measurements is common in order to increase the measurement redundancy and the stability of DSSE algorithms [11], [21]–[23], [26], [27], [30], [34], [35], [37], [40]–[44]. Since measurement data is generally corrupted by noise, an SE problem is required to be capable of refining the correct states from the corrupted measurements [1], [38], [45]. The measurement platform must be also designed such that the SE algorithm remains stable against failures of or bad data injection by some measurement units [38], [45]. It is typical to consider single-contingency in PMU-placement problem. Criticality detection [45] is the method used in the literature to define the critical measurement signal the failure of which makes the SE algorithm unstable. By introducing $\mu$PMUs [5], installation of larger number of measurement units will be less costly in the future. However, observability analysis is expected to be still the interest of the research groups working on SE problems to create effective, less expensive measurement platforms.

In this project, both PSSE and DSSE problems are formulated based SDP. Observability analysis and criticality detection corresponding to SDP formulation is performed to avoid any instability of the algorithm in case of measurement noise, bad data, or failure of measurement units. The algorithms are implemented in MATLAB and solved using SDPA solver. Several simulations are carried out, and the proposed methods are tested on multiples IEEE test networks.

## Scope of the work

1- The conventional iterative LSE method is not guaranteed to seek the globally optimal solution of the SE problem due to the non-convexity of the AC system. In contrary, SDP formulation of the SE problem is promising to find the globally optimal solution, but some rank relaxations should be considered to make the solution set convex.
2- A DSSE which captures the unbalanced nature of the distribution networks is modeled by a reduced Y matrix eliminating all nodes not actually existing in the system. The proposed method is capable of handling single-phase, three-phase, and multiphase power systems.
3- In active distribution networks, there are unbalanced loads and DERs which must be considered in DSSE problem. Measurement units on appropriate places address the imbalance of DERs such as DG, ESS, and loads. The measurements corresponding to these resources can be easily fed to the proposed SDP-based DSSE.
4- Observability analysis is inevitable part of any SE algorithm. Although it is quite similar in SDP formulation and other techniques, some other techniques must be performed to help SDP-based SE converge to the correct solution. This method is sensitive to the set of measurements for all networks and to the level of noise in large networks. These issues and their solutions are addressed later in this report.
5- Bad data detection/identification methods proposed for Newton's method does not work for SDP-based SE. A method is proposed to detect bad data suspects before running the SE algorithm and identifying the real bad measurements.

# Semidefinite Programming based State Estimation

## Power System State Estimation (PSSE)

PSSE is a technique to estimate the state of power systems (voltage magnitudes and angles of buses) from a limited number of measurements in order to have an observation on how the power system is being operated. PSSE is a complex optimization problem due to the nonlinear functions relating measurements and voltage states. It is also prone to convergence issues and locally optimal solutions because of its high non-convexity.

PSSE problem is based on the following measurement model:

$$z_i = h_i(X) + u_i \tag{1}$$

where $z_i$ denotes the $i^{th}$ measurement such as active and reactive powers flowing in transmission line, voltage magnitude, and net injected active and reactive powers to the nodes. $h_i$ is a nonlinear function describing $z_i$ in terms of the state variable vector $X$, and $u_i$ denotes the additive measurement noise to the $i^{th}$ measurement assumed to be a zero-mean independent Gaussian random variable. That is, $u_i \sim N(0, \sigma_i^2)$ such that $\sigma$ is the standard deviation corresponding with the noise of $i^{th}$ measurement device. Obviously, the expected value for i-th measurement can be expressed as $E(z_i) = h_i(X)$.

According to the independence of measurement noises, we have:

$$p(U = 0) = \prod_{i=1}^{m} p(u_i = 0) \tag{2}$$

The objective of PSSE is to seek the vector $X$ maximizing likelihood function $f_U(0) = f_Z(E(Z))$ which is equivalent to the probability of $E(Z) = Z$ or $U = 0$:

$$X = \text{argmax} \quad f_U(0) = \prod_{i=1}^{m} f_{u_i}(0) = \prod_{i=1}^{m} f_{u_i}(0) = \prod_{i=1}^{m} \frac{1}{\sqrt{2\pi\sigma_i}} e^{-\frac{(z_i - E(z_i))^2}{\sigma_i^2}} \tag{3}$$

s.t. $E(z_i) = h_i(X) \quad \forall_{i=1,\cdots,m}$

In order to simplify the optimization procedure, the function is commonly replaced by its logarithm to determine the optimum parameter values as follows:

$$X = \text{argmax} \quad \prod_{i=1}^{m} \frac{1}{\sqrt{2\pi\sigma_i}} e^{-\frac{(z_i - E(z_i))^2}{\sigma^2}} = \text{argmin} \sum_{i=1}^{m} \frac{(z_i - E(z_i))^2}{\sigma_i^2} \tag{4}$$

s.t. $E(z_i) = h_i(z_i) \quad \forall_{i=1,\cdots,m}$

Defining $W_{ii} = \frac{1}{\sigma_i^2}$ and $r_i = z_i - E(z_i)$, the optimization problem will look like:

$$X = \text{argmin} \sum_{i=1}^{m} W_{ii} \cdot r_i^2 \tag{5}$$

s.t. $E(z_i) = h_i(z_i)$, $W_{ii} = \frac{1}{\sigma_i^2}$, $r_i = z_i - E(z_i)$

The solution of above optimization problem is called WLS estimator for X.

Newton's method is the most common algorithm to solve WLS problem. However, this method is subject to local optima because of the non-convex nature of the SE problem. SDP-based SE is the alternative algorithm which transforms the SE problem to a convex format and seeks the global optimal solution.

## SDP-based PSSE Formulation

The PSSE problem can be formulated in SDP format based on the SDP formulation of power system equations presented in [32] and [16]. Employing the SDP representation of power system equations, the measurements can be expressed by the matrix $W = UU^T$ where $U$ is the vector of real and imaginary values of voltage phasors with the size of $2n$ for an n-bus system.

Denoting $e_1, e_2, \cdots, e_n$ as the standard basis vectors in $\mathbf{R}^n$, let us define a number of matrices for every node $k$, and any branch $lm$ between the sending node $l$ and receiving node $m$. These matrices will be called SDP matrices hereafter in this paper.

$$Y_k := e_k e_k^T Y_{bus} \tag{6}$$

$$Y_{lm} := (y_{lm}^* + y_{lm}) e_l e_l^T - y_{lm} e_l e_m^T, \quad y_{lm} = Y_{bus}(l, m) \tag{7}$$

$$\mathbf{Y}_k := \frac{1}{2} \begin{bmatrix} Re\{Y_k + Y_k^T\} & Im\{Y_k^T - Y_k\} \\ Im\{Y_k - Y_k^T\} & Re\{Y_k + Y_k^T\} \end{bmatrix} \tag{8}$$

$$\overline{\mathbf{Y}_k} := \frac{-1}{2} \begin{bmatrix} Im\{Y_k + Y_k^T\} & Re\{Y_k - Y_k^T\} \\ Re\{Y_k^T - Y_k\} & Im\{Y_k + Y_k^T\} \end{bmatrix} \tag{9}$$

$$\mathbf{Y}_{lm} := \frac{1}{2} \begin{bmatrix} Re\{Y_{lm} + Y_{lm}^T\} & Im\{Y_{lm}^T - Y_{lm}\} \\ Im\{Y_{lm} - Y_{lm}^T\} & Re\{Y_{lm} + Y_{lm}^T\} \end{bmatrix} \tag{9}$$

$$\overline{\mathbf{Y}_{lm}} := \frac{-1}{2} \begin{bmatrix} Im\{Y_{lm} + Y_{lm}^T\} & Re\{Y_{lm} - Y_{lm}^T\} \\ Re\{Y_{lm}^T - Y_{lm}\} & Im\{Y_{lm} + Y_{lm}^T\} \end{bmatrix} \tag{10}$$

$$M_k := \begin{bmatrix} e_k e_k^T & 0 \\ 0 & e_k e_k^T \end{bmatrix} \tag{11}$$

$$X := [Re\{V\}^T \quad Im\{V\}^T]^T \tag{12}$$

$$W := XX^T \tag{13}$$

According to the matrices defined in (6)-(13), common parameters measured in power system can be defined as below:

$P_{inj,k} = Tr\{Y_k W\}$: Net active power injection from bus k into the grid

$Q_{inj,k} = Tr\{\overline{Y_k} W\}$: Net reactive power injection from bus k into the grid

$P_{lm} = Tr\{Y_{lm} W\}$: Active power drawn by bus l from branch l-m (direction is defined based on [32])

$Q_{lm} = Tr\{\overline{Y_{lm}} W\}$: Reactive power drawn by bus l from branch l-m

$|V_k|^2 = Tr\{MW\}$: Voltage magnitude on bus k

The measurement $z_i$ can be explained in SDP format below, where $A_i$ is the SDP matrix corresponding to measurement $z_i$.

$$z_i = Tr(A_i W) + u_i \tag{14}$$

Thus, the PSSE optimization problem can be expressed as below:

$$X = \text{argmin} \quad \sum_{i=1}^{m} \frac{(z_i - Tr(A_i W))^2}{\sigma_i^2} \tag{15}$$

s.t. $\quad W = XX^T$

where $A_i$ is the SDP matrix corresponding to measurement $z_i$.

Definition of matrix $W$ necessitates its rank to be 1. Therefore the optimization problem above can be replaced by:

$$W = \text{argmin} \quad \sum_{i=1}^{m} \alpha_i \tag{16}$$

s.t. $\quad \alpha_i \geq \frac{(z_i - Tr(A_i W))^2}{\sigma_i^2} \quad \forall_{i=1,\cdots,m}$

$\quad W \geq 0$

$\quad \text{rank}(W) = 1$

The first constraint also needs to be expressed in SDP format. $\alpha_i \geq \frac{(z_i - Tr(A_i W))^2}{\sigma_i^2}$ leads to $\alpha_i \sigma_i^2 - (z_i - Tr(A_i W))^2 \geq 0$ which is guaranteed if $\begin{bmatrix} \alpha_i \sigma_i^2 & z_i - Tr(A_i W) \\ z_i - Tr(A_i W) & 1 \end{bmatrix}$ is a positive semidefinite matrix. This problem is still a non-convex optimization problem due to the rank constraint. By relaxing this constraint the optimization problem is converted to a convex SDP problem. This, the final form of the relaxed SE optimization problem is as follows:

W= argmin $\sum_{i=1}^{m} \alpha_i$ \hfill (17)

s.t. $\quad \begin{bmatrix} \alpha_i \sigma_i^2 & z_i - Tr(A_i W) \\ z_i - Tr(A_i W) & 1 \end{bmatrix} \geq 0 \quad \forall_{i=1,\cdots,m}$

$$W \geq 0$$

## SDP-based DSSE:

In PSSE problem, a single-phase model of the power network is formulated. Although this approach is common and reasonable to analyze transmission networks because, it is not capable of addressing the unbalanced nature of distribution networks. In DSSE problem, therefore, the network must be modeled in three-phase format. In a distribution feeder it is also possible for a bus to be connected to any combination of three phases (a, b, and c). These distribution networks are called multiphase and must be analyzed in a multiphase format to decrease computation expenses.

The advantage of SDP-based SE is to deal with SDP matrices which are derived directly from the network Y-matrix. Therefore, having Y-matrix is sufficient to derive SDP matrices for SE problem formation. Fortunately, Y-bus of a multiphase distribution system can be derived by following the rules of forming Y-matrix of a transmission system. For an N-bus AC transmission network, the size of Y-matrix is $N \times N$ while it is $3N \times 3N$ for a three-phase N-bus distribution feeder. For a multiphase distribution system, the size of Y-matrix is $N' \times N'$ where $N' = \sum_{i=1}^{N} d_i$ and $d_i$ denotes the number of phases connected to bus $i$.

Using the Y-matrix of distribution network, one can derive SDP matrices for SDP-DSSE problem. The other procedures are the same as those explained in SDP-PSSE section.

# Observability Analysis, Noise Effect Reduction, and Bad Data Detection/Elimination

Observability analysis and criticality detections of SDP-based SE are quite similar to that of the Newton's method, which is comprehensively investigated in the literature [1, 40, 49]. However, there are some differences between SDP-based SE and other methods. In the methods other than SDP, the number of state variables is equal to twice as many as nodes ($2N'$), whereas the number of state variables in SDP is equal to $N'(2N'+1)$ because of removing the rank constraint and symmetry of matrix $W$. Therefore, the observability of SDP-based SE is much more vulnerable than the other methods for same measurement platforms. In this section, the reasons of observability issues of SDP-based SE will be studied. The vulnerability of SDP method versus lack of information about the line power flow signals on both sides has been addressed and the solution developed is proposed. The sensitivity of large power networks to noise corruption and it solution employing μPMU is also discussed. Further, bad data detection and elimination is discussed later.

## Sources of Redundancy in SDP format

In linear algebra, one set of $m$ equations in terms of $n$ unknowns has a unique answer if there are exactly $n$ linearly independent equations among the $m$ equations. In the SE methods other than SDP, the number of variables is exactly the number of states to be defined, so the number of linearly independent equations corresponding to measurement set is a reliable signal to define whether the network is observable.

In SDP-based SE algorithm, however, the number of variables is $N(2N+1)$ for an $N$-node, $M$-branch network. Some of these variables have no effect on the final answer because they do not have any non-zero coefficients, and some are the same due to symmetry of the matrix $W$. Removing such float variables, which encompasses $W(i,j)$ elements such that there is no line connecting bus $i$ to bus $j$, leads to $3N+4M$ distinct variables (3 variables for each node, 4 variables for each line). On the other hand, the biggest possible size of measurement set (including active and reactive powers injected to buses or flowing through the lines as well as bus voltage magnitudes plus one reference phase) is $3N+4M$, but there are only $N+2M$ linearly independent equation in such a measurement set. Therefore, the number of measurements is not enough to guarantee the uniqueness of the final solution regardless of the rank constraint.

For example, the number of state variables in the state estimation problem for the redial 41-bus network presented in IEEE Standard 399 [46] is 3403 and can be reduced down to $3n + 4m = 283$ by removing the float state variables and symmetry of the matrix $W$. The maximum number of measurement signals is 283 including 121 linearly independent signals.

There are some reasons generating linear dependence of the measurements. First, the Kirchhoff's Current Law (KCL) imposes linear independence between net injection power (active and reactive) at each node and that flowing in the lines connected to the same bus. That is,

$$P_{inj,l} = -\sum_{m=1}^{n} P_{lm} \Rightarrow Y_l = -\sum_{m=1}^{n} Y_{lm} \tag{18}$$

$$Q_{inj,l} = -\sum_{m=1}^{n} Q_{lm} \Rightarrow \overline{Y_l} = -\sum_{m=1}^{n} \overline{Y_{lm}} \tag{19}$$

The other redundancy comes from the voltage magnitude of adjacent nodes and the power flow on the connecting lines. According to the definition of $y_{lm}$, $Y_{lm}$, $\overline{Y_{lm}}$, $M_l$, and $M_m$ the following equations stand:

$$Im(y_{lm}) \times (Y_{lm} + Y_{ml}) + Re(y_{lm}) \times (\overline{Y_{lm}} + \overline{Y_{ml}}) = 0 \tag{20}$$

$$Re(y_{lm}) \times (Y_{lm} - Y_{ml}) + Im(y_{lm}) \times (\overline{Y_{lm}} - \overline{Y_{ml}}) = |y_{lm}|^2 \times (M_l - M_m) \tag{21}$$

Now, it is obvious why there are 162 linearly dependent equations $(2 * 41 + 2 * 40)$ in the example of IEEE Standard 399.

To summarize:

1- There are $3n + 4m$ distinct state variables in SDP formulation of SE
2- There are at most $3n + 4m$ measurement signal capable to be fed to SE problem
3- There are two points of redundancy at each node for active and reactive power injections
4- There are two points of redundancy for each line relating active and reactive power flows on the line and the voltage magnitudes of the ends
5- From all possible measurements, only $(3n + 4m) - (2n - 2m) = n + 2m$ equations are linearly dependent which is lesser than the number of state variables
6- If rank constraint is not considered, the solution of the SE problem is not necessarily unique.

## Vulnerability of SDP-SE and Solution

In the observability analysis of a power system for state estimation purpose, it is typical to consider active and reactive measurements as a pair. Also, it is assumed that having power flow on one of the ends of a line is enough as the power flow on that specific line. However, this assumption does not work with SDP-based SE due to elimination of some linearly independent equations from measurement set. In fact, when there is only power measurement from one side of the line, the system gets unobservable and SDP is not able to seek the right answer. Two ideas can be discussed as solutions to this issue.

First option is to install measurement devices on both sides of a line being monitored, which mandates installation of a large volume of measurement units in the network. This solution cannot be considered as the best, or even a good, solution due to its expenses.

Second option is based on estimating the actual active and reactive power signals of the other side of the line from the data already gathered by measurement units and feeding them into the state estimation problem. For this purpose, some solutions are proposed below assuming that $P_{ij}$ and $Q_{ij}$ are measured but $P_{ji}$ and $Q_{ji}$ are not.

It must be noted that one of the parameters of $P_{ji}$ and $Q_{ji}$ is enough to recover the observability because adding the other parameters not only cannot improve the observability of the system due to the linear dependence of $P_{ij}$, $Q_{ij}$, $P_{ji}$, and $Q_{ji}$ described in (3-3), but also brings some inaccuracy to the problem.

1. The X/R ration in transmission lines is large, and even in most cases, R=0. Therefore, the active power loss on the line is approximately zero i.e. $P_{ji} \approx -P_{ij}$. If $r_{ij} = 0$, the variance corresponding to $P_{ji}$ can be set equal to that of $P_{ij}$. Otherwise, it should be big enough[1] to reflect the inaccuracy of the value of $P_{ji}$. This technique can be also used for distribution networks, but bigger value for variance may be needed. In some cases where $x_{ij} = 0$, it is better to recover the observability by introducing $Q_{ji} \approx -Q_{ij}$ instead.
2. The other method is to apply the power loss on the line into the approximation process. The power loss percentage can be derived from the historical efficiency data of the entire power system being studied. Therefore, $P_{ji}$ can be derived by the following equation if $(1 - \eta) * 100\%$ of the power is dissipated on the line:

$$P_{ij} + P_{ji} + P_{loss_{ij}} = 0 \Rightarrow P_{ji} = \begin{cases} -P_{ij}/\eta & P_{ij} \geq 0 \\ -\eta P_{ij} & P_{ij} < 0 \end{cases} \qquad (22)$$

The variance corresponding to $P_{ji}$ must be selected big enough in the SE problem.

3. Since $P_{ij}$ and $Q_{ij}$ signals are measured and the network parameters are known, the value of $P_{ji}$ or $Q_{ij}$ can be derived analytically as well:

$$P_{ij} + P_{ji} + P_{loss_{ij}} = 0 \Rightarrow P_{ji} = -P_{ij} - r_{ij}I_{ij}^2 = -P_{ij} - r_{ij}\frac{S_{ij}^2}{|V_i|} == -P_{ij} - r_{ij}\frac{P_{ij}^2+Q_{ij}^2}{|V_i|^2} \quad (23)$$

$$Q_{ij} + Q_{ji} + Q_{loss_{ij}} = 0 \Rightarrow Q_{ji} = -Q_{ij} - x_{ij}I_{ij}^2 = -Q_{ij} - x_{ij}\frac{S_{ij}^2}{|V_i|} == -Q_{ij} - x_{ij}\frac{P_{ij}^2+Q_{ij}^2}{|V_i|^2} \quad (24)$$

---

[1] 1000 times as high as variance of power on the other side is selected as a big enough value for variance in this work.

If $|V_i|$ is not measured, the assumption of $|V_i| = 1$ seems reasonable, and a big value for variance reflects any inaccuracy in the approximation. In this case, using the latest value of $|V_i|$ is the other option.

If $|V_i|$ is also measured, its value can be easily used to derive better approximation of $P_{ji}$ or $Q_{ji}$. Although a big value for variance reflects the approximation inaccuracy, it is desired to limit the variance signal to the value which is as large as required. In this case, where all signals are measured, a probabilistic analysis helps to define a reasonable variance value. Variance analysis of the equation (3-6) leads to:

$$var(P_{ji}) = var(P_{ij}) + r_{ij}^2 E\left(cov(P_{ij}^2, Q_{ij}^2)\right) var\left(\frac{1}{|V_i|^2}\right) + r_{ij}^2[var(P_{ij}^2) + var(Q_{ij}^2) + cov(P_{ij}^2, Q_{ij}^2)]E\left(\frac{1}{|V_i|^2}\right) + r_{ij}^2[var(P_{ij}^2) + var(Q_{ij}^2) + cov(P_{ij}^2, Q_{ij}^2)]var\left(\frac{1}{|V_i|^2}\right) \quad (25)$$

where $var(X) = \sigma_X^2$. An upper limit for $var(P_{ji})$ defines a reasonable value to be fed into the SE problem.

Since the noise is of the variables $P_{ij}$ and $Q_{ij}$ are assumed to be white noise, $P_{ij}^2$ and $Q_{ij}^2$ can be expressed as follows:

$$P_{ij} = E(P_{ij}) + \sigma_{P_{ij}} Y \Rightarrow P_{ij}^2 = \mu_{P_{ij}}^2 + 2E(P_{ij})\sigma_{P_{ij}} + \sigma_{P_{ij}}^2 Y^2 \quad (26)$$

where $\mu_{P_{ij}} = E(P_{ij})$ and Y denotes a random variable following a standard normal probability distribution. $Y^2$ follows a Chi-squared distribution, so $var(Y^2) = 2$. Since $prob\left(P_{ij} \geq \mu_{P_{ij}} - 3\sigma_{P_{ij}}\right) = prob\left(P_{ij} \leq \mu_{P_{ij}} + 3\sigma_{P_{ij}}\right) \geq 0.99$, the following analyses are reasonable.

$$E(P_{ij}^2) \leq \left(|P_{ij}| + 3\sigma_{P_{ij}}\right)^2 \quad (27)$$

$$E(Q_{ij}^2) \leq \left(|Q_{ij}| + 3\sigma_{Q_{ij}}\right)^2 \quad (28)$$

$$E\left(\frac{1}{|V_i|^2}\right) \leq \frac{1}{\left(|V_i| - 3\sigma_{|V_i|}\right)^2} \quad (29)$$

$$var(P_{ij}^2) = 2\sigma_{P_{ij}}^4 + 4\sigma_{P_{ij}}^2 \mu_{P_{ij}}^2 \leq 2\sigma_{P_{ij}}^4 + 4\sigma_{P_{ij}}^2 \left(|P_{ij}| + 3\sigma_{P_{ij}}\right) \quad (30)$$

$$var(Q_{ij}^2) = 2\sigma_{Q_{ij}}^4 + 4\sigma_{Q_{ij}}^2 \mu_{P_{ij}}^2 \leq 2\sigma_{Q_{ij}}^4 + 4\sigma_{Q_{ij}}^2 \left(|Q_{ij}| + 3\sigma_{Q_{ij}}\right) \quad (31)$$

$$var\left(\frac{1}{|V_i|^2}\right) = E\left(\frac{1}{|V_i|^4}\right) - E\left(\frac{1}{|V_i|^2}\right)^2 \leq \frac{1}{\left(|V_i| - 3\sigma_{|V_i|}\right)^4} - \frac{1}{\left(|V_i| + 3\sigma_{|V_i|}\right)^4} \quad (32)$$

$$cov(P_{ij}^2, Q_{ij}^2) = E(P_{ij}^2, Q_{ij}^2) - E(P_{ij}^2)E(P_{ij}^2) \leq \left(|P_{ij}| + 3\sigma_{P_{ij}}\right)^2 \left(Q_{ij} + 3\sigma_{Q_{ij}}\right)^2 -$$
$$\left(|P_{ij}| - 3\sigma_{P_{ij}}\right)^2 \left(Q_{ij} - 3\sigma_{Q_{ij}}\right)^2 \tag{33}$$

Applying the upper limits provided in (27)-(3-33) into (25) leads to an efficient (as large as required) value for $var(P_{ji})$ to be used as input in the SE algorithm. If the algorithm utilizes the signal of $Q_{ji}$ instead, same logic can be used to derive the value of $var(Q_{ji})$.

## Noise Effect Reduction

According to the observability analysis described above, there are $3n + 4m$ distinct state variables in the SDP-based SE while at most $n + 2m$ linearly independent measurements may be gathered from the network. Then, the SE results are very sensitive to the measurement signals and their level of noise, especially in multiphase radial distribution networks due to large number of nodes and less redundancy. As an example, there are 117 nodes and 457 lines in IEEE 37-bus test feeder shown in Fig 1, which introduces 2179 unknown against with only 1023 linearly independent equations. However, SDP-based DSSE estimates the voltage states perfectly when there is no noise interfering measurements, but adding different levels of noise to the measurements reveals the sensitivity of the solutions to the level of noise. Fig 2 shows the simulation results of SDP-based DSSE for IEEE 37-bus test feeder at different levels of noise presented in Table I. As seen, the algorithm perfectly estimates the voltage states with no noise, but the deviation of its results from the correct answer increases by the level of noise. Thus, large distribution networks are more vulnerable against high noise interference.

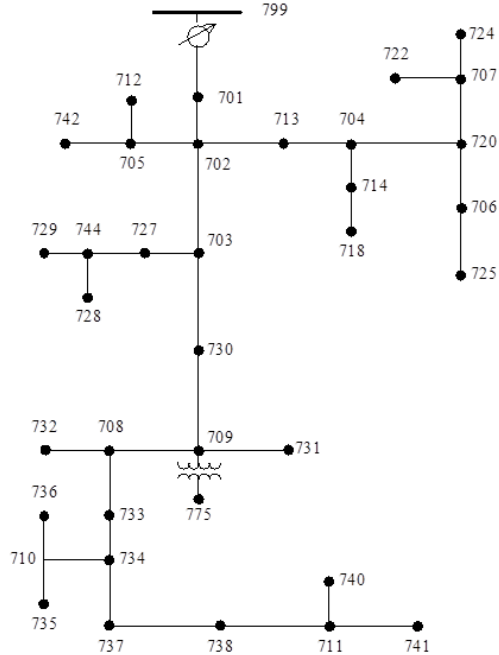

Fig. 1- IEEE 37-bus distribution test feeder

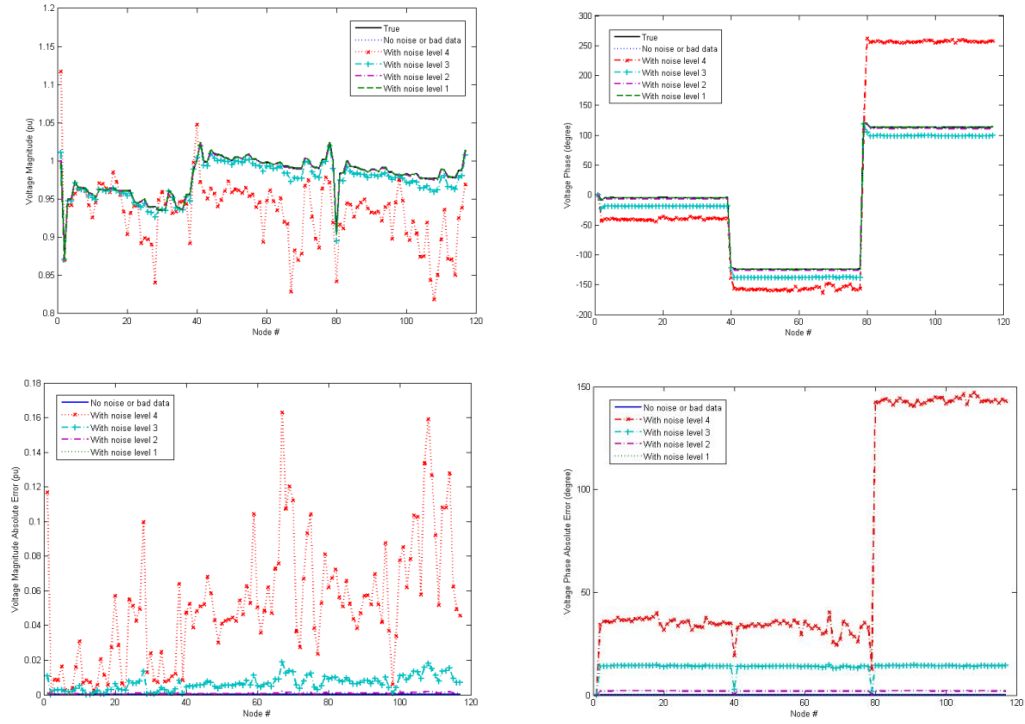

Fig. 2- SDP-based DSSE results of IEEE 37-bus test feeder for various levels of noise

Table I - Different levels of Noise

|  | Level 0 | Level 1 | Level 2 | Level 3 | Level 4 |
|---|---|---|---|---|---|
| Active Power Injection Noise | 0 | $1.5 \times 10^{-5}$ | $1.5 \times 10^{-4}$ | $1.5 \times 10^{-3}$ | $1.5 \times 10^{-2}$ |
| Reactive Power Injection Noise | 0 | $1.5 \times 10^{-5}$ | $1.5 \times 10^{-4}$ | $1.5 \times 10^{-3}$ | $1.5 \times 10^{-2}$ |
| Voltage Magnitude Noise | 0 | $1 \times 10^{-5}$ | $1 \times 10^{-4}$ | $1 \times 10^{-3}$ | $1 \times 10^{-2}$ |
| Active Line Power Noise | 0 | $2 \times 10^{-5}$ | $2 \times 10^{-4}$ | $2 \times 10^{-3}$ | $2 \times 10^{-2}$ |
| Reactive Line Power Noise | 0 | $2 \times 10^{-5}$ | $2 \times 10^{-4}$ | $2 \times 10^{-3}$ | $2 \times 10^{-2}$ |

On the other hand, smaller distribution networks are more resilient against noise. Fig. 3 illustrates the SDP-based DSSE results for levels 0 and 4 of noise applied to IEEE 13-bus test feeder which has 38 nodes and 107 lines. The results demonstrate the toleration of the algorithm for smaller distribution networks. This fact along with the advent of $\mu$PMU devices [28] introduces an idea to solve the SE problem more reliably.

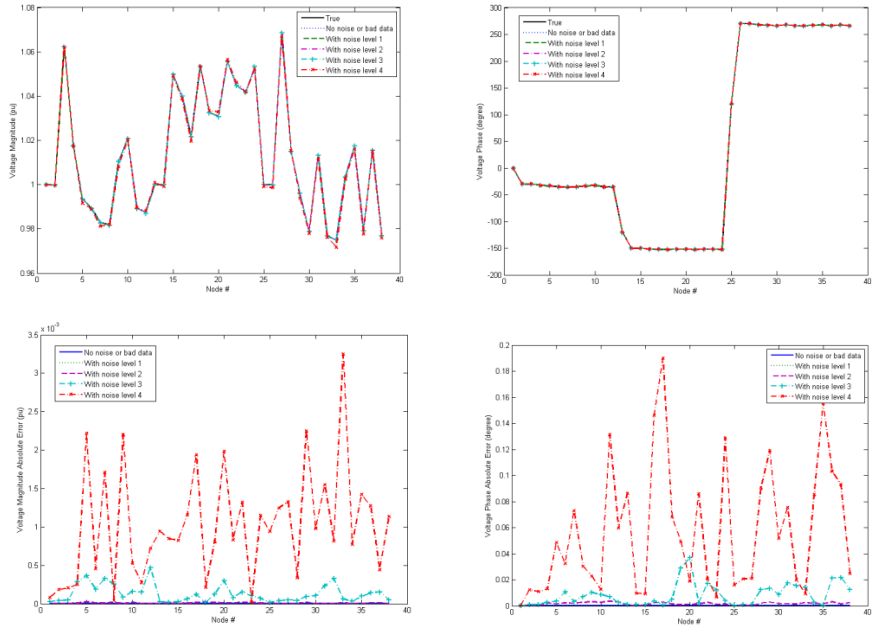

Fig. 3- SDP-based DSSE results of IEEE 13-bus test feeder for various levels of noise

**Decoupled SDP-based SE:** A large, observable network can be separated into several sub-networks. Each integrated sub-network is observable and includes some nodes and some lines connecting the included nodes. If each sub-network has at least one *μ*PMU which measures the corresponding voltage phasor respect to a global reference point, the voltage states of each sub-network can be estimated independently from the other sub-networks. The algorithm is capable to solve the SDP-SE problem even if *μ*PMU devices are shared between several sub-networks. It must be also noted that some lines, called tie-lines hereafter, are connected between the sub-networks through the boundary nodes. The measurements corresponding to tie-lines may be neglected in SE procedure or may be used to update power injection of the boundary nodes. The final solutions of sub-networks form the final solution of the entire network.

For example, IEEE 37-bus test feeder can be separated as illustrated in Fig. 4. Each sub-network forms a tree, and the lines 704-713 and 709-730 are tie-lines. There should be at least one μPMU in each sub-network. It is assumed that there are two *μ*PMU devices on the buses 713 and 730. Each μPMU is only connected to one of the phases (nodes) of the corresponding bus. The SE problem of each sub-network is solved via SDP formulation assuming that the angles are zero on *μ*PMU-equipped nodes. The final solution of entire network is obtained by shifting the voltage phasors of each sub-network as much as the phase difference of its *μ*PMU and the reference.

The results, which are illustrated in Fig. 5, demonstrate that the decoupled SDP-based SE perfectly mitigates the effects of noise in huge power networks.

In addition to the improvement in noise reduction, decoupled SDP-based SE is capable to be carried on in a parallel way in order to accelerate the state estimation process.

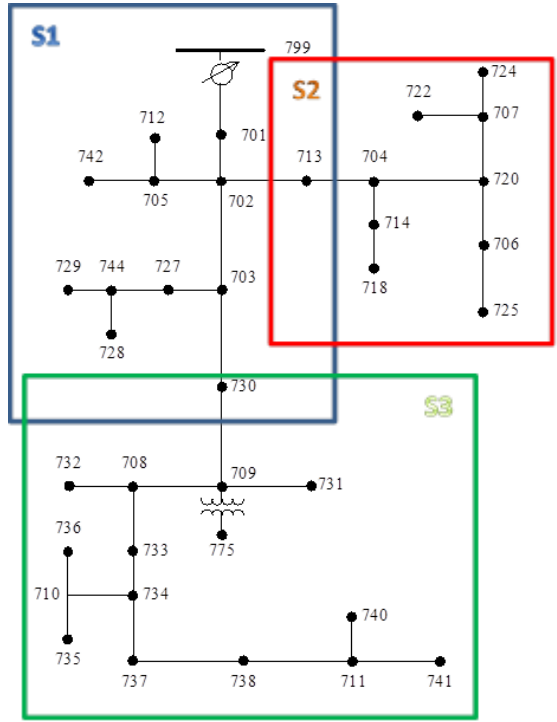

Fig. 4- Definition of sub-networks for IEEE 37-bust test feeder for decoupled SDP-DSSE

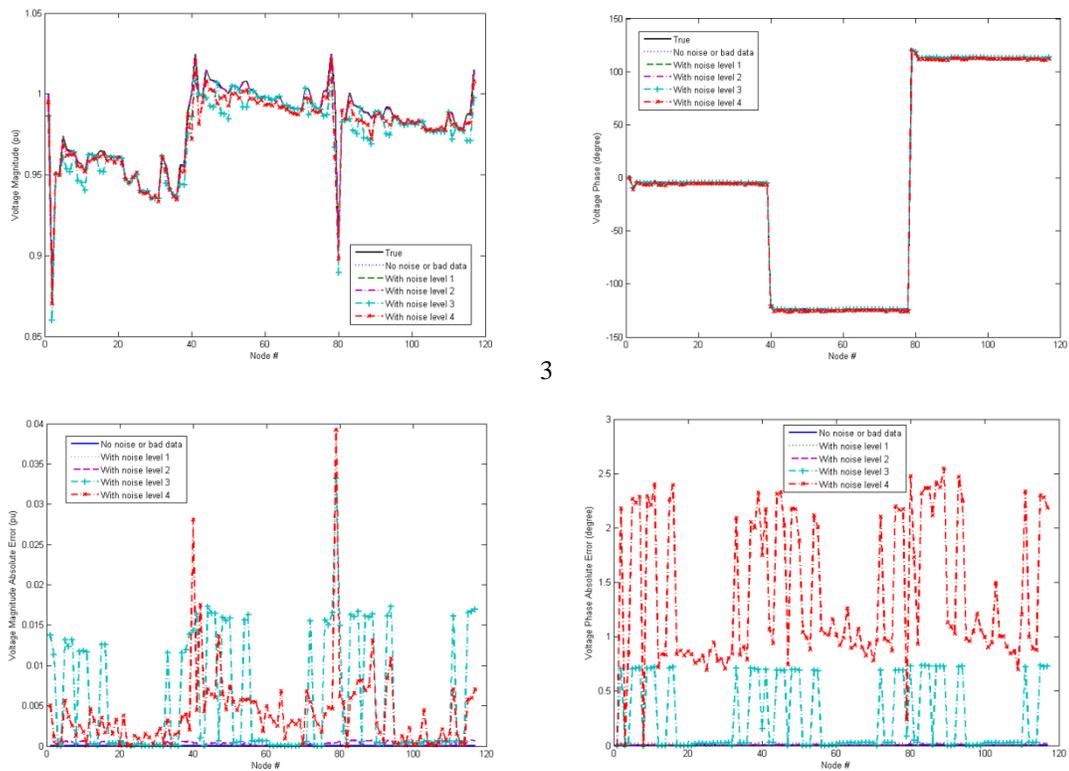

Fig. 5- The results of decoupled SDP- DSSE for IEEE 37-bus test feeder for various levels of noise

**Topology Detection and Network Separation:** Given that the entire network is observable, any integrated sub-network is observable as well. Then decoupled SDP-PSSE algorithm is capable to solve the state estimation algorithm for huge power networks. Since huge networks (e.g. IEEE 8500-node test feeder) cannot be divided into sub-networks manually, two algorithms has been developed to first identify the topology of the network (topology detection) and then to allocate the buses (or nodes) to integrated, ideally-sized sub-networks for decoupled SE algorithm. Dividing the network to sub-networks defines some tie-lines whose power flow measurements may be ignored in the SE problem or may be used to update the net power injection of the boundary nodes.

Some assumptions and explanation are required for better understanding of the algorithm. Without loss of generality, we assume the network is multiphase radial. Each bus $i$ has up to three phases forming the set $N_i$. The buses $i$ and $j$ are called adjacent if any of their phases (nodes) are connected together. Among the adjacent buses of bus $i$, the closest one to the main feeder is called the parent of bus $i$ ($P_i$) and the others form its children set $C_i$. The generation set $G_i$ of a bus is recursively defined as the set of the children of bus $i$ and their generations (children, grandchildren, grand-grandchildren, …) $G_i = C_i \cup (\cup_{j \in C_i} G_j)$. The size of $G_i$ is called rank of bus $i$ hereafter. The ancestor set of bus $\Lambda_i$ also has a recursive definition of the set of the parent of bus $i$ as well as its ancestors $\Lambda_i = \{P_i\} \cup \Lambda_{P_i}$.

The topology detection, which is presented in ALGORITHM 1, is a recursive algorithm to define the generation set of all buses in the network. It starts from the main feeder, finds its adjacent buses, and refines its children from them. Line data and admittance matrix may be employed to identify the adjacent buses. Then, the algorithm is repeated for each child until the entire network is processed.

ALGORITHM 1: Topology Detection

1- Start from the main feeder ($i = 1$).
2- Identify $N_i$ and all the nodes connected to any node $k \in N_i$ based on Y-matrix ($Y_{kl} \neq 0 \Leftrightarrow l \in M_i$). Put them in the set $M_i$.
3- Define the adjacent set $C_i = \{j : M_i \cap N_j \neq \emptyset, j \neq i, j \notin \Lambda_i\}$.
4- For any bus $c$ in $C_i$, if b has not been processed yet:
   a. Let $C_i = C_i \cup \{c\}$, $P_c = i$, and $\Lambda_c = \{i\} \cup \Lambda_i$.
   b. Let $i = c$ and go to 2.
   c. $G_{P_i} = G_{P_i} \cup \{i\} \cup G_i$.
5- For any bus $i$ of the network: $R_i = |G_i|$.

Since the topology detection algorithm does not start processing of any buses which are already processed, it can be applied to meshed network as well. Although the final results are corresponding to a radial tree of the meshed network, they are still valid for separation process.

ALGORITHM 2 proposes the separation procedure. First, a desirable value is selected for the size of sub-networks. The generation set of the bus with closest rank to desirable value forms a new sub-network. All the buses assigned to the sub-network just created are removed from the network, and the algorithm continues until all buses are allocated to sub-networks.

ALGORITHM 2: Network Separation

1- Select the desired value for the number of buses $(d)$ in each sub-network and let $k = 1$.
2- As long as there is at least one bus with a rank more than zero (while $\max(R_i) \neq 0$):
   a. Find the bus $i$ with the closest rank to $d$ ($i = \text{argmin}|d - R_i|$).
   b. Form a new sub-network $S_k = \{i\} \cup G_i$.
   c. Let $G_a = G_a - G_i$ and $R_a = R_a - R_i$ for any bus $a \in \Lambda_i$.
   d. Let $G_s = \emptyset$ and $R_s = 0$ for any bus $s \in S_k$.
   e. Let $k = k + 1$

A posterior investigation about the sub-networks created and their common buses determines the candidates for $\mu$PMU installation.

## Bad Data Detection

Measurements are always subject to errors due to different reasons such as finite accuracy of the meters and telecommunication media. Large errors also happen when meters have biases, drifts, and wrong connections or telecommunication fails or suffers noises from unexpected interferences. Apart from these, the approximation technique introduced in this project to improve the observability of the system may inject some bad data into state estimator.

Then, it is of the significant functionalities of a state estimator to detect the measurement errors, also known as bad data, to identify the source of error, and to eliminate it from measurement set until the source of error exist. This function requires the system not only to be observable, but also to have a sufficient level of redundancy among measurements.

Some bad data such as negative voltage magnitude are obvious and can be identified and removed from the measurement set before running the state estimation. Other errors need some posterior investigations after the state estimation results are defined.

Different methods for bad data detection and identification have been discussed thoroughly in [1]. A brief review of the bad data detection/identification technique for Newton's method is provided below:

1. Calculate the Jacobian matrix H relating measurement differences to the state variable differences ($\Delta Z = H \Delta X + U$).
2. Derive the residual sensitivity matrix $S = I - (H^T R^{-1} H)^{-1} H^T R^{-1}$ where R is the diagonal covariance matrix of the measurement noises.
3. Calculate residual values corresponding to all measurements. ($r_i = \Delta z_i - \Delta \hat{z}_i \quad \forall i$)
4. Calculate normalized residuals $\left(r_i^N = \frac{|r_i|}{\sqrt{R_{ii} S_{ii}}} \quad \forall i\right)$ the probability distributions of which are expected to follow a standard normal distribution
5. In order to decide on the existence of bad data, the largest element in $r_i^N$ is compared against a statistical threshold which is chosen according to the desired level of detection sensitivity.
6. If it is revealed that bad data exists, the measurement corresponding to the largest normalized residual is removed and SE is performed again.

Unlike Newton's method, there is no Jacobian matrix in SDP-based SE. Instead, every measurement is related to the state variables in matrix $W$ according to (14). If $A_i^j$ and $W^j$ denotes the j-th column of $A_i$ and $W$ respectively, (14) can be rewritten as:

$$z_i = A_i^{'} \times W^{'} + u_i \tag{34}$$

where $A_i' = [A_i^1, A_i^2, \cdots, A_i^{2n}]^T$, and $W' = [W^1, W^2, \cdots, W^{2n}]^T$. Therefore, the following equation defines the linear relationship between measurement vector $Z$ and the state vector $W'$ where $Z = [z_1, z_2, \cdots, z_m]^T$, $A' = [A_1', A_2', \cdots, A_m']^T$, and $U = [u_1, u_2, \cdots, u_m]^T$:

$$Z = A' \times W' + U \tag{35}$$

Regarding (2-17), there is a linear relationship between measurement sets and state variables in SDP format. As mentioned before, the matrix W has some float elements which have no effects on the results. Also, some equations rows of equations in (2-17) are exactly same due to symmetry of the matrices $W$ and $A_i$'s. Eliminating the float elements from matrix $W'$ and combining the symmetric elements of $W'$ leads to a reduced version of state variables $W'_r$. Likewise, a reduced matrix $A'_r$ is obtained by removing zero columns of matrix $A'$, which are corresponding to float elements of $W$. Therefore, (2-17) can be refined as below:

$$Z = A_r^{'} \times W_r^{'} + U \tag{36}$$

The algorithm explained to detect and eliminate bad data for Newton's method may be modified for SDP method as below:

1. Calculate the Jacobian matrix $A'_r$ relating measurements to the state variables ($Z = A'_r W_r + U$).
2. Derive the residual sensitivity matrix $S = I - \left(A'^T_r R^{-1} A'_r\right)^{-1} A'^T_r R^{-1}$ where R is the diagonal covariance matrix of the measurement noises.
3. Calculate residual values corresponding to all measurements. ($r_i = z_i - \hat{z}_i \quad \forall i$)

4. Calculate normalized residuals $\left(r_i^N = \frac{|r_i|}{\sqrt{R_{ii}S_{ii}}} \quad \forall i\right)$ the probability distributions of which are expected to follow a standard normal distribution
5. In order to decide on the existence of bad data, the largest element in $r_i^N$ is compared against a statistical threshold which is chosen according to the desired level of detection sensitivity.
6. If it is revealed that bad data exists, the measurement corresponding to the largest normalized residual is removed and SE is performed again.

The matrix R is diagonal and the size of the matrix $A'_r$ is $m \times (3N + 4M)$ and $m \leq (3N + 4M)$. Also, the rank of matrix $A'$ is at most $N + 2M$, which causes the matrix $A'^T R^{-1} A'$ to have a size of $(3N + 4M) \times (3N + 4M)$ with a maximum rank of $N + 2M$ which is less than $3N + 4M$. Consequently, the matrix $A'^T_r R^{-1} A'_r$ is a singular matrix and it cannot be inversed to obtain derive the matrix $S$. Thus, the algorithm cannot be applied to SDP-based SE.

There is another option to detect and identify single bad data using information provided about the sources of redundancy. For example, from (3-1), $u_{P_l} = P_{inj,l} + \sum_{m=1}^n P_{lm} \; \forall l$ is expected to follow a zero-mean normal distribution with a variance equal to $\sigma^2 = var(P_{inj,l}) + \sum_{m=1}^n var(P_{lm})$. If all the signals involved are in the measurement set, $u_{P_l}$ is compared against a statistical threshold defined by the desired level of detection sensitivity. If it is revealed that bad data exists, all signals involved are considered as bad data suspects.

The complete procedure is described as follows:

1. For each node $l$, calculate $u_{P_l} = P_{inj,l} + \sum_{m=1}^n P_{lm}$ and $u_{Q_l} = Q_{inj,l} + \sum_{m=1}^n Q_{lm}$ if all signals involved are measured. The set of such signals is called redundant set of node $l$.

2. For each node $l$, compare $u_{P_l}$ and $u_{Q_l}$ versus a statistical threshold to define whether it follows a zero-mean standard distribution with $\sigma_{P_l}^2 = var(P_{inj,l}) + \sum_{m=1}^n var(P_{lm})$ and $\sigma_{Q_l}^2 = var(Q_{inj,l}) + \sum_{m=1}^n var(Q_{lm})$ respectively. If bad data is detected, save all measurements as suspects in the suspect set $Z_l^s$.

3. For each branch $l - m$, calculate $u_{1_{lm}} = Im(y_{lm}) \times (P_{lm} + P_{ml}) + Re(y_{lm}) \times (Q_{lm} + Q_{ml})$ and $u_{2_{lm}} = Re(y_{lm}) \times (P_{lm} - P_{ml}) + Im(y_{lm}) \times (Q_{lm} - Q_{ml}) - |y_{lm}|^2 \times (|V_l|^2 - |V_m|^2)$ if all signals involved are measured. The set of such signals is called redundant measurement set of branch $l - m$.

4. For each branch $l - m$, compare $u_{1_{lm}}$ and $u_{2_{lm}}$ versus a statistical threshold to define whether it follows a zero-mean standard distribution with $\sigma_{1_{lm}}^2 = Im(y_{lm})^2 \times var(P_{lm} + P_{ml}) + Re(y_{lm})^2 \times var(Q_{lm} + Q_{ml})$ and $\sigma_{2_{lm}}^2 = Re(y_{lm})^2 \times var(P_{lm} - P_{ml}) + Im(y_{lm})^2 \times var(Q_{lm} - Q_{ml}) - |y_{lm}|^4 \times var(|V_l|^2 - |V_m|^2)$. If bad data is detected, save all measurements as suspect set $Z_{lm}^s$.

5. If there is no suspect for bad data, perform SDP-SE once and release the results as final results. End. Otherwise, go to step 6.

**Bad data identification:**

6. There is at least one set of suspects; for all combinations where one measurement $z_i^S \in Z_i^S$ is assumed bad data from each suspect set $Z_i^S$: a) for each measurement selected, calculate its value and variance based on other measurements in its own suspect set $\{z_k^S = f(z_{S\setminus k}^S), var(z_k^S) = g(var(z_{S\setminus k}^S))\}$, b) perform SDP-SE using the calculated values, c) regenerate the measurement signals from the SE results and compare to the originally measured values.

7. Release the results of the case with the least error as the final result. End.

# Simulations

The algorithms proposed are evaluated on different Test systems. The radial 41-bus test system of IEEE Standard 399 [46], IEEE 30-bus test system [47], and IEEE 39-bus test system [48] are selected for SDP-PSSE. Multiphase IEEE 13-bus, three-phase IEEE 37-bus and multiphase IEEE 123- bus test feeders [49] are also considered to study SDP-DSSE algorithm. For each test system, three different cases (without noise or bad data, with noise, with bad data) are simulated. Table I illustrates the standard deviation of white noise signals to emulate the noise-corrupted measurements as suggested in [12]. In all cases the line power flow signals are provided from one of the ends. So, to overcome the vulnerability of the SDP algorithm versus lack of the line power flow measurements, which is addressed in section 3, the active power value of the other side $(P_{ji})$ and its standard deviation $(\sigma_{P_{ji}})$ have been assumed to be equal to $-P_{ij}$ and $1000 * \sigma_{P_{ij}}$, respectively. Big value of $\sigma_{P_{ji}}$ addresses the possible error of the approximation applied.

Table II – Noise Data

| Measurement type | Standard deviation |
|---|---|
| Active power injection | 0.015 pu |
| Reactive power injection | 0.015 pu |
| Active power flow | 0.02 pu |
| Reactive power flow | 0.02 pu |
| Voltage magnitude | 0.01 pu |

To test the capability of the algorithm to detect and identify bad data, one of the measurements is considered to be corrupted by large error while the others are corrupted only by noise. The measurement signals for PSSE and DSSE problems are derived from MATPOWER [50] and OpenDSS [51] respectively, and the final results are benchmarked versus the load flow results obtained from these software packages.

## Test System 1: SDP-PSSE for IEEE Std 399 (Radial 41-bus system)

For the first numerical study, the SDP-based PSSE is tested against the radial 41-bus test system introduced in IEEE Standard 399 as shown in Fig. 6. The voltage and power measurement units are illustrated on the network. The signals of active and reactive power flowing through all lines are measured from one side (closer to the meter) to the other side. There are four voltage meters, two of them are connected to the generation buses and the other two are connected to load buses. The power injection on the generation buses are also measured to create some redundancy in the measurement set. Although, the power injection signals corresponding to zero-injection buses may be used as additional measurements, they have been neglected in this case study. The data failure in this test system is also assumed to happen in the power injection measurement of bus

50, which results in a bad data suspect set including power flows of lines 50-51 and 50-3 as well as power injection on bus 50.

Fig. 7 depicts the simulation results for all cases, which demonstrate the accuracy of the SDP-based PSSE to handle both noise and bad data. In spite of lack of information about the power flow data on the other sides of the lines and applying the approximation technique, the voltage magnitude and phase of the case with no noise or bad data perfectly lies on the expected (true) values. For the other two cases, the largest error of the voltage magnitude is 0.005 pu and that of the voltage phase is 0.08 degree.

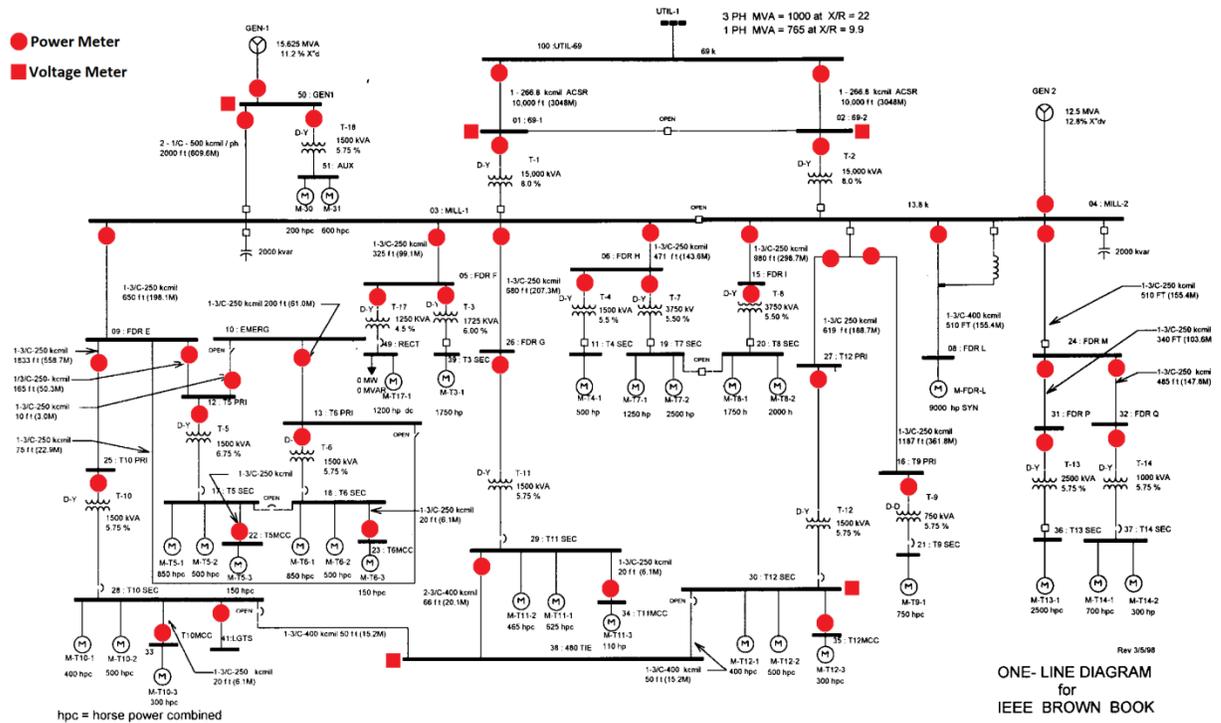

Fig. 6- IEEE Standard 399 41-bus test feeder with measurement devices

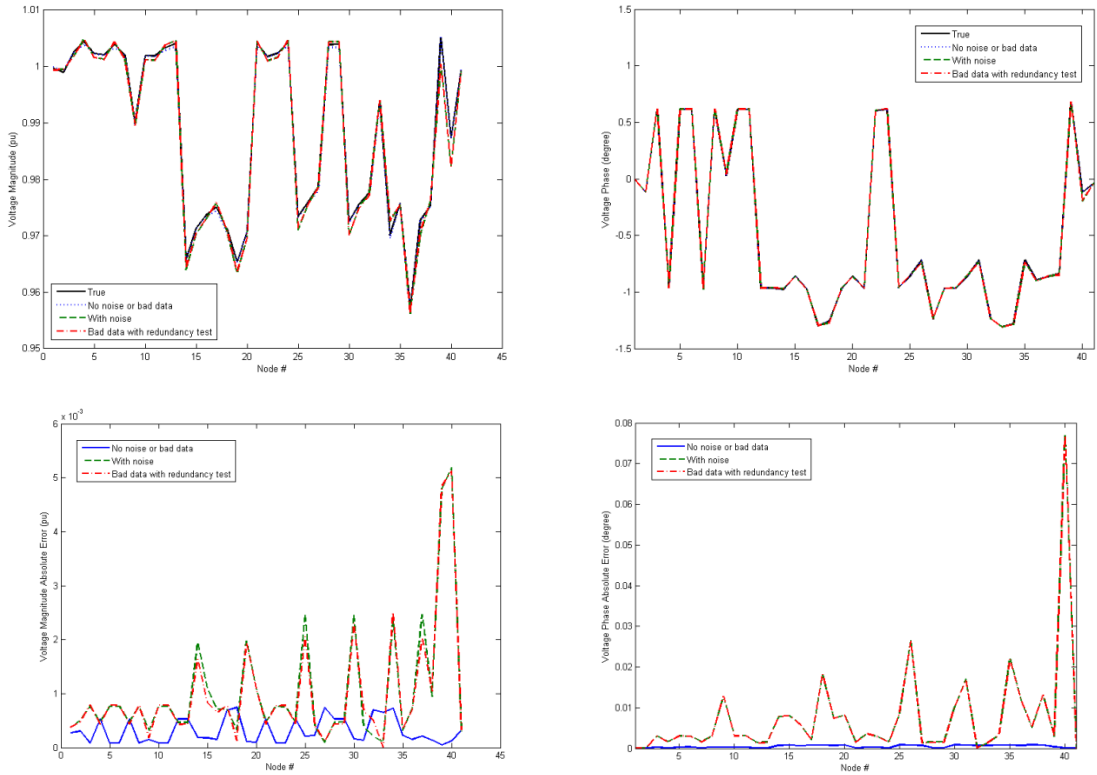

Fig. 7- Simulation results of SDP-based PSSE for IEEE Std. 399 41-bus test feeder for different cases

## Test System 2: SDP-PSSE for IEEE 30-bus test system

Unlike the IEEE 399 Standard test feeder, the IEEE 30-bus test network is a meshed network which is studied as the next test system. Although the state estimation of a meshed network using SDP formulation is more concerned about the eliminated rank constraint, it is provided by more measurements due to more redundancy of meshed networks compared to radial ones. For this test system, active and reactive power signals of the transmission lines are measured from one of their sides. The power injections of the generation buses as well as their voltage magnitude are the other measurement signals fed to SE problem. Fig. 8 illustrates the topology of the measurement devices considered on IEEE 30-bust test network. In this numerical example, the bad data is assumed to happen in the signal of power flowing from bus 1 toward bus 2. Therefore, the bad data suspects are power flows of lines 1-2 and 1-2 as well as power injection of bus 1.

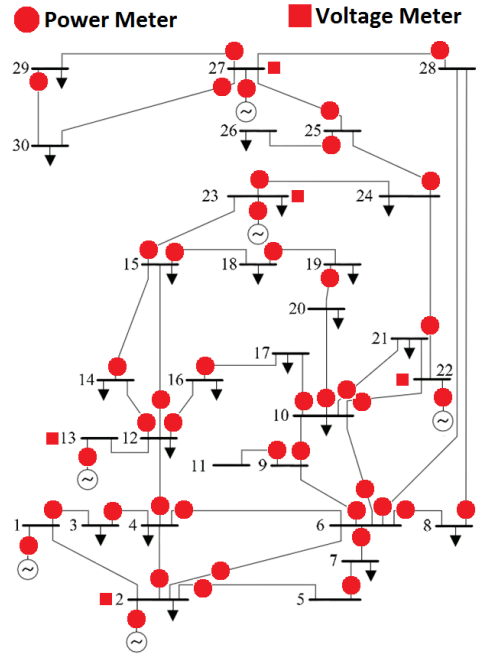

Fig. 8- IEEE 30-bus test system with measurement devices

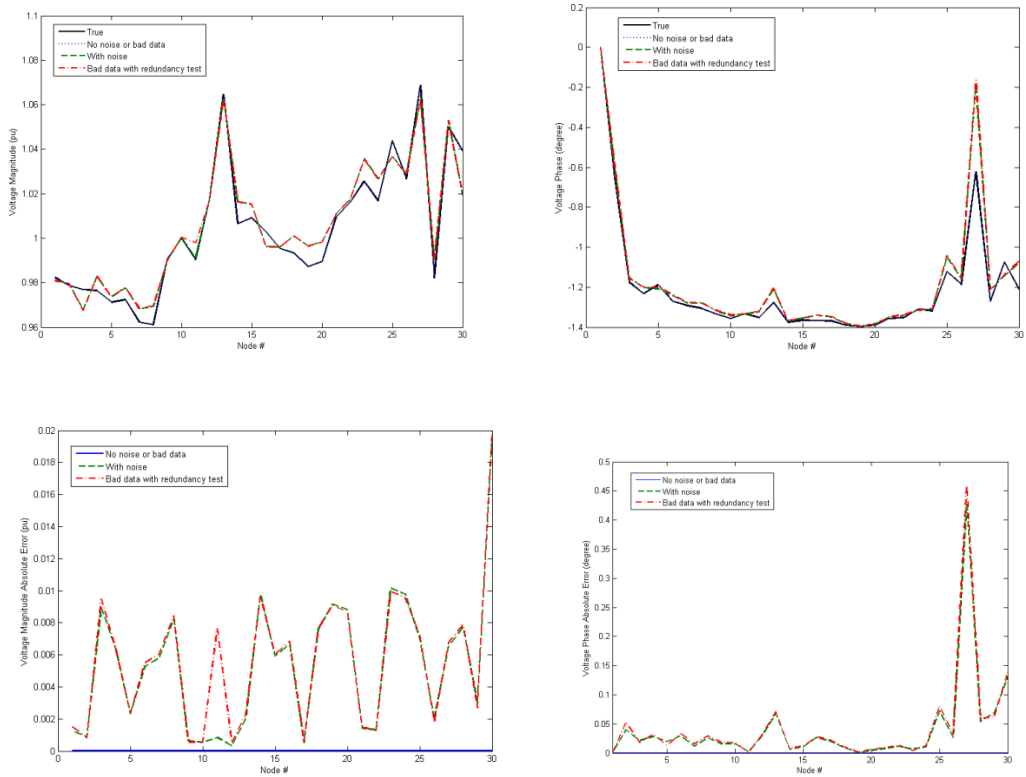

Fig. 7- Simulation results of SDP-based PSSE for IEEE 30-bus test system for different cases

As shown in Fig. 9, the results of the case with no noise have almost zero deviation from the expected values which demonstrates the accuracy of the SDP-PSSE algorithm and the approximation procedure to recover line power flow signals. The noise corruption of the measurement signals leads to deviations of voltage magnitudes and angles whose maximum values are 0.02 pu and 0.45 degree, respectively. Same deviations observed in the case of bad data, which is also corrupted by the same noise signals, demonstrates the efficiency of the single bad data detection/identification proposed in the section 3.

## Test System 3: SDP-PSSE for IEEE 39-bus test system

Fig. 10 illustrates the IEEE 39-bus test network and applied measurement devices. The power flows of all transmission, as well as power injection and voltage magnitude on the generation buses. There are twelve zero-injection buses (1, 2, 5, 6, 7, 9, 11, 12, 14, 17, 19, and 22) in the network whose net power injections are considered in the SE problem due to less redundancy in distribution networks. These buses along with generation buses form the redundancy points of the measurement configuration designed. Bad data is presumed to occur on the power signal of the line 12-13 which causes line power flows 11-12 and net injection of bus 12 to be detected as the other bad data suspects.

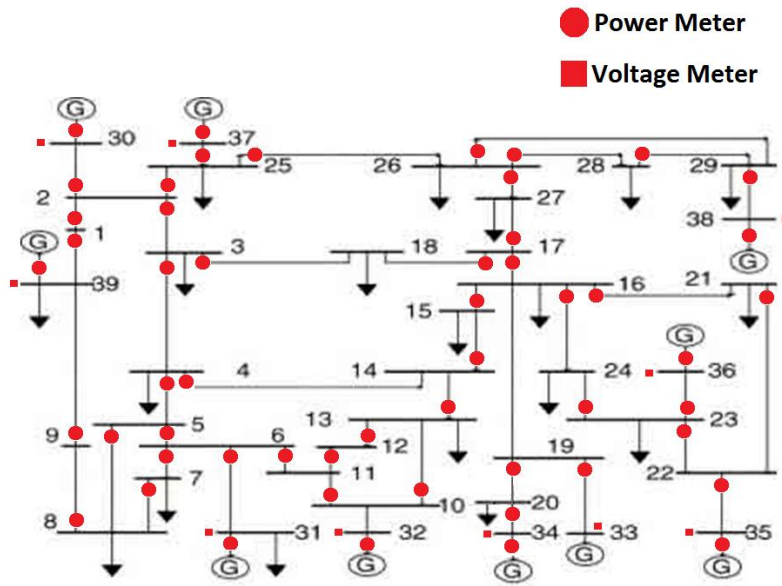

Fig. 10- IEEE 39-bus test system with measurement devices

The results corresponding to this test system are depicted in Fig. 11. The SDP-based PSSE solves the case without noise accurately with the results matching the expected values. The cases considering noise corruption and bad data shows some deviations as low as 0.016 pu in voltage magnitudes and 0.09 degrees in voltage angles.

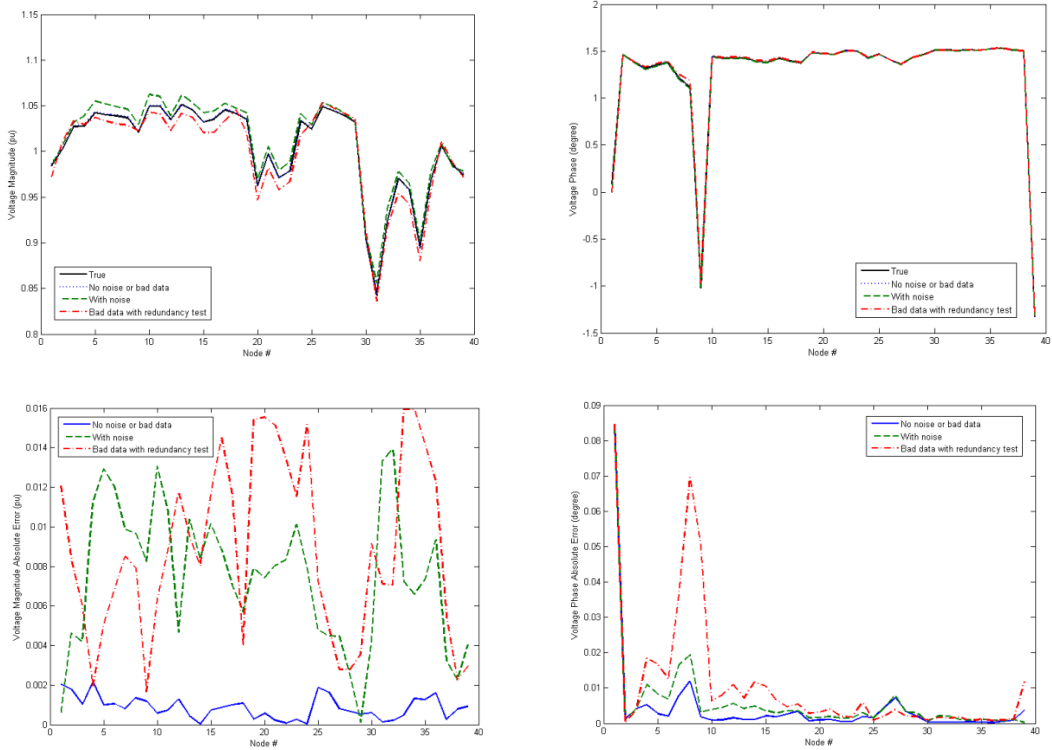

Fig. 11- Simulation results of SDP-based PSSE for IEEE 39-bus test system for different cases

## Test System 4: SDP-DSSE for multiphase IEEE 13-bus test feeder (38-node)

In order to investigate the functionality of SDP-based DSSE, first, it is tested against three-phase IEEE 13-bus test feeder. Since the buses in this feeder have various numbers of phases, it is counted as a multiphase distribution network. The switch between buses 671 and 692 is closed in this study. In this case, the measurement set encompasses line power flows, power received from the upper network, and three voltage magnitude signals. Fig. 12 illustrates the test feeder as well as the configuration of the measurement units. One bad data occurs on the phase A of bus 650.

Fig. 13 depicts the results which are sorted based on phases (A, B, and C) for more clear illustrations. The results show that the estimated states match the expected values perfectly in zero-noise case. The voltage magnitude and phase deviations of all nodes are less than .007 pu and 1.08 degrees, respectively, which proves low sensitivity of the SDP-DSSE to noise interference for this test system.

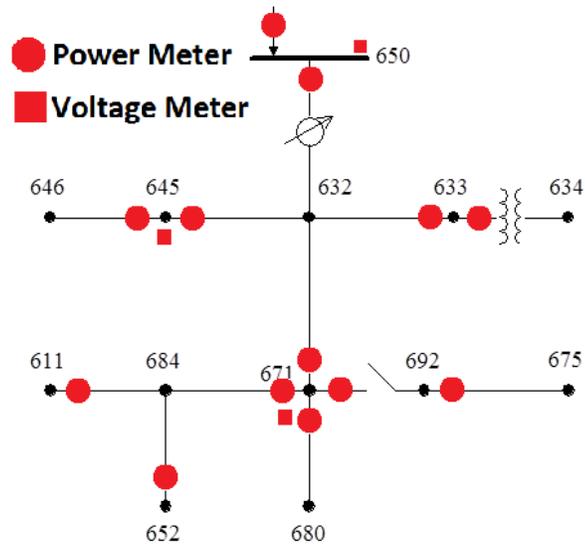

Fig. 12- IEEE 13-bus test system with measurement devices

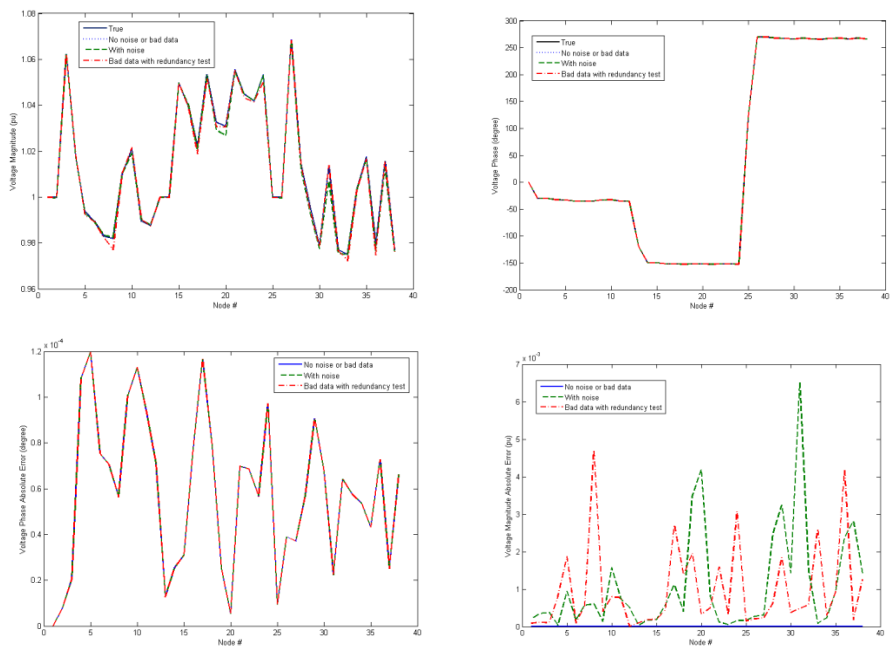

Fig. 13- Simulation results of SDP-based DSSE for IEEE 13-bus test feeder for different cases

## Test System 5: SDP-DSSE for three-phase IEEE 37-bus test system (117-node)

Fig. 4 shows the IEEE 37-bus test feeder which is separated to three sub-networks. Each sub-network forms a tree, and the lines 704-713 and 709-730 are tie-lines. There should be at least one μPMU in each sub-network. It is assumed that there are two μPMU devices on the buses 713 and 730. Each μPMU is only connected to one of the phases (nodes) of the corresponding bus. The SE problem of each sub-network is solved via SDP formulation assuming that the angles are zero on μPMU-equipped nodes. The final solution of entire network is obtained by shifting the voltage phasors of each sub-network as much as the phase difference of its μPMU and the reference. The results which are illustrated in Fig. 5 demonstrate that the decoupled SDP-based SE perfectly mitigates the effects of noise in huge power networks.

## Test System 6: SDP-DSSE for multiphase IEEE 123-bus test system (278-node)

In order to study the effectivity of the algorithm versus larger networks, the multiphase IEEE 123-bus test feeder is studied. This system, which encompasses 278 nodes, has several switches which are normally-open or normally-closed. In the first simulation the status of the switches are assumed to be constant i.e. always closed or always open. Therefore, they do not need to be considered in decoupling process of the network into sub-networks. As seen in Fig. 14, the network is divided into 8 sub-networks illustrated in different colors. The phase A of buses 13, 76, and 300 are equipped with μPMU which are shown by a diamond. The μPMU on is shared on the bus 13 is shared between sub-networks 1, 2, and 3; the one on bus 76 is common between sub-networks 6, 7, and 8 while the last one which is on the bus 300 belongs to both sub-networks 4 and 5. There are two tie-lines, one between sub-networks 5 and 6, and the other one between 3 and 6. Since the switch between sub-networks 3 and 8 is normally open, it is not considered as tie-line in this simulation. The data gathered for tie-lines might be either ignored in the SE process or used to update the information corresponding to the boundary buses 54, 57, 67, and 97.

Fig. 15 depicts the results of decoupled SDP-based DSSE for the cases of zero-noise, with noise, and with bad data/noise, considering the abovementioned decoupling format. The maximum deviation of voltage magnitude in zero-noise case is as low as .004 pu while it is about five times greater in the other two cases. The deviations of the voltage angles do not exceed 0.1 degree in zero-noise case and 0.7 degree in the other ones. These results strongly demonstrate the efficiency of the decoupled SDP-based DSSE to handle state estimation of a huge power network.

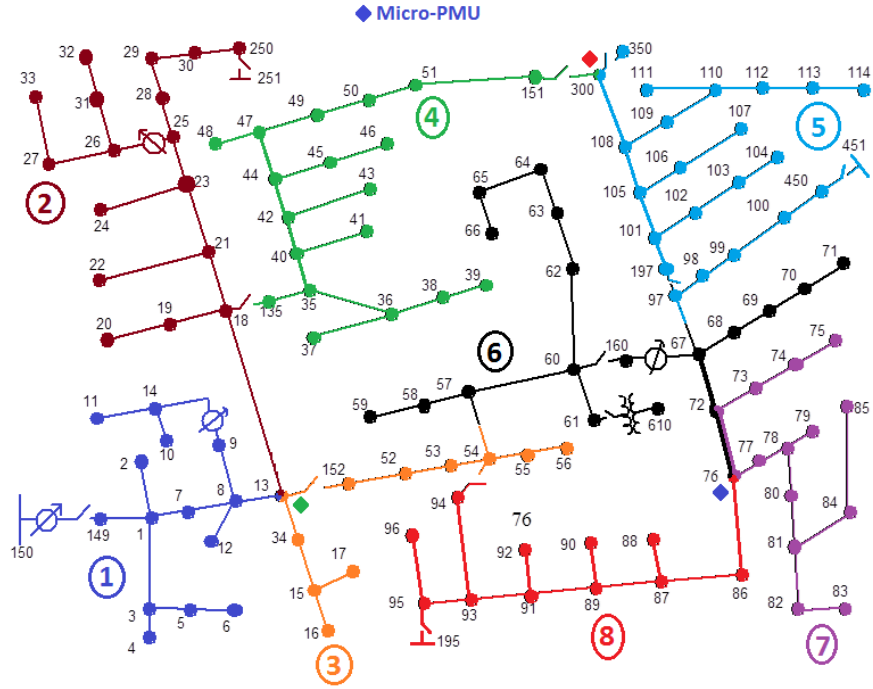

Fig. 14- IEEE 123-bus test system and sub-networks definition 1

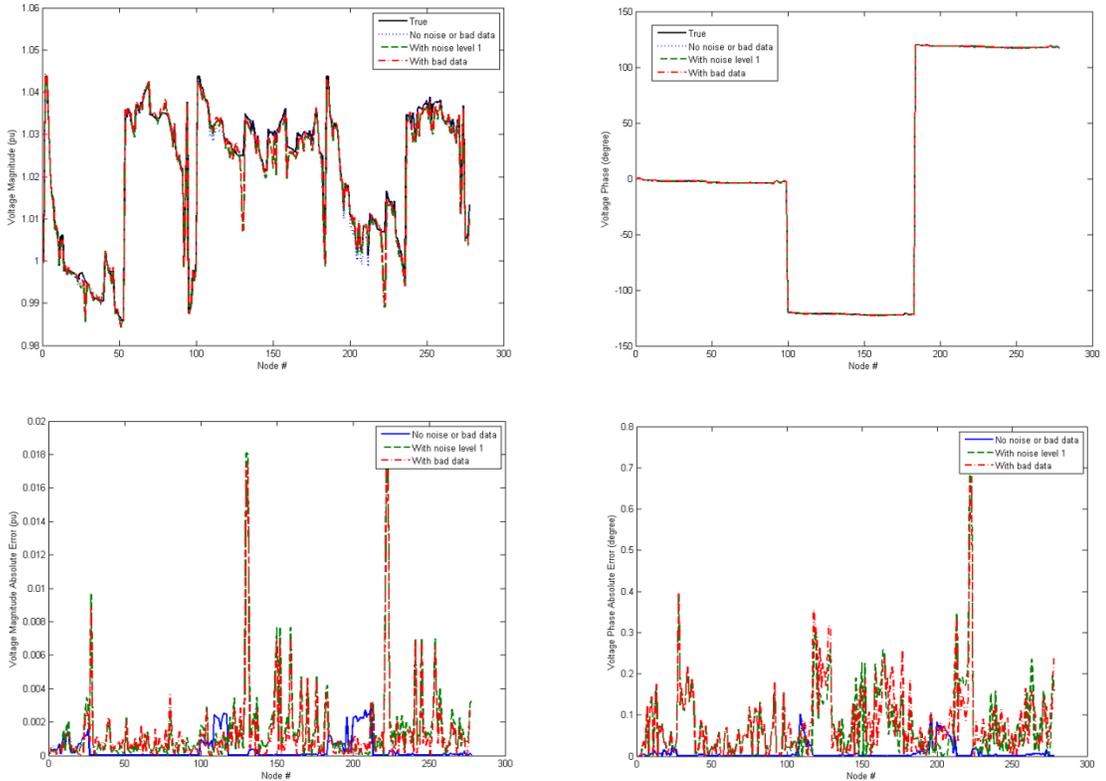

Fig. 13- Simulation results of decoupled SDP-based DSSE for sub-network definition 1 of IEEE 123-bus test feeder for different cases

Although the algorithm of decoupled SDP-DSSE was carried on successfully in the previous simulation, the configuration of switches is required to be considered in µPMU placement and network decoupling process since the switches can be on or off any time. Therefore, the µPMU devices must be allocated such that the every sub-network is equipped by at least one µPMU for any configuration of the switches. One of the possible decoupling formats is to divide the network into sub-networks from the points where switches are connected, i.e. switches will be considered as tie-lines. Fig. 27 shows the new five sub-networks defined based on switches in the network. As the switches are not always connected it is not possible to share µPMU devices between sub-networks. Hence, each µPMU is dedicated to only one sub-network and might be connected to any node of the sub-network.

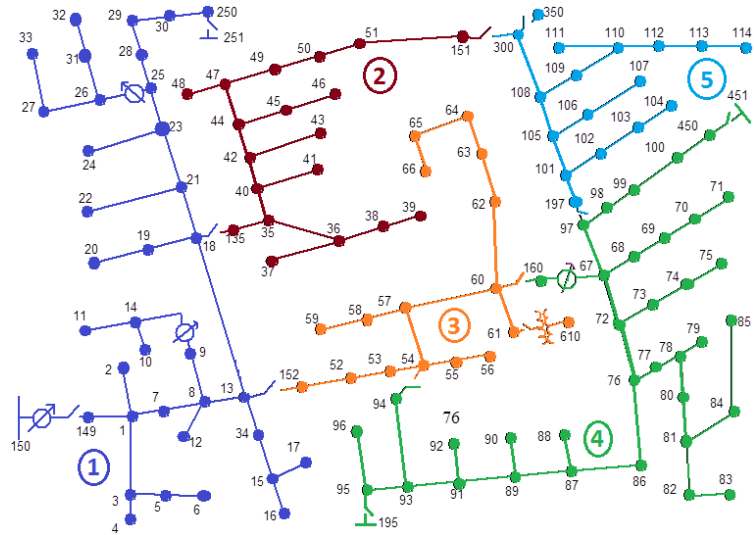

Fig. 16- IEEE 123-bus test system and sub-networks definition 2

Fig. 28 illustrates the results corresponding to decoupled SDP-DSSE for the IEEE 123-bus test feeder whose sub-networks are based on Fig. 28. The results, i.e. voltage magnitude deviation less than 0.02 pu and voltage angle deviation as low as 0.8 degree, demonstrate that the algorithm mitigates the adverse effects of noise interference and data failure in the large networks.

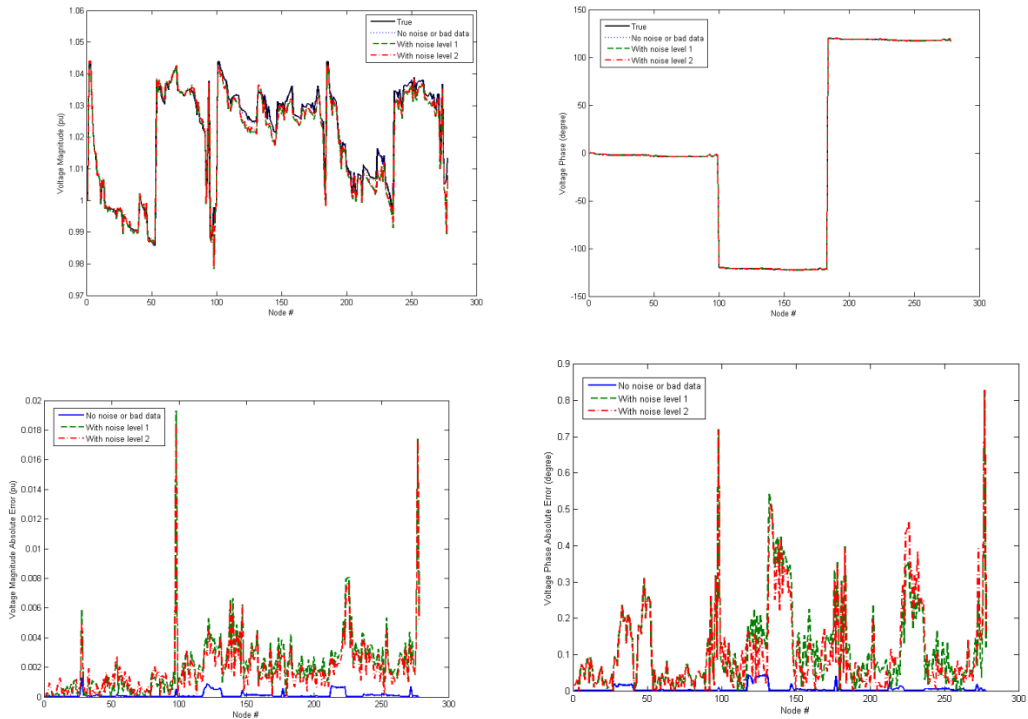

Fig. 15- Simulation results of decoupled SDP-based DSSE for sub-network definition 2 of IEEE 123-bus test feeder for different cases

## Test System 7: SDP-DSSE for multiphase EPRI Circuit-5 (2998-bus) test feeder

To demonstrate the flexibility of the algorithm, it has been tested against EPRI Circuit-5 test system. This test system is a multiphase 2998-bus network encompassing 3437 nodes. Of these nodes, 1149 nodes are phase A, 1152 are phase B, and 1136 nodes are phase C.

The rank detection and network separation algorithms divide the network to 50 sub-network with an average size of 60 buses. Then, the decoupled SDP-algorithm is carried on for zero-noise and several levels of noise (2, 3, 4) defined in Table I.

Due to the large number of buses in this case, statistical results are more appropriate to be reported than the simulation results themselves. Table III presents the statistical indices of the simulation results. These results demonstrate the capability of the decoupled SDP-SE algorithm to mitigate the effect of noise in such a huge system. For the zero-case noise, the average deviations of voltage magnitude and phase are less than 0.0005 pu and 0.05 degree respectively. Adding noise at level 2 triples the deviation of voltage magnitude (0.0012) and doubles that of voltage phase (0.09). Apparently, higher levels of noise generate more deviation in state values sought by SE algorithm versus their true values. The simulation results for the case with noise at level 3 shows mean deviations less than 0.005 pu and 0.55 degree in voltage magnitude and phase respectively. Remarkable deviations in the case with a noise at level 4 necessitate running the separation algorithm with a lower desired size for sub-networks, which results in higher number of sub-networks.

Table I – Statistics of simulation results of EPRI Circuit 5 test feeder

| Parameter | Index | No noise | Level 2 | Level 3 | Level 4 |
|---|---|---|---|---|---|
| Voltage Magnitude Error | Root-Mean-Square (RMS) | 0.001201 | 0.003469 | 0.01341 | 0.09388 |
| | Average | 0.000421 | 0.001158 | 0.005327 | 0.035791 |
| | Maximum | 0.008404 | 0.021716 | 0.363432 | 0.971411 |
| Voltage Phase Error | Root-Mean-Square (RMS) | 0.109614 | 0.224764 | 1.142754 | 6.272216 |
| | Average | 0.040339 | 0.088148 | 0.539509 | 3.588595 |
| | Maximum | 0.713487 | 1.17768 | 20.20768 | 42.33768 |

Along with the abstract indices proposed in Table I, Fig.16 depicts the histogram of the absolute errors of voltage magnitudes of the entire network for different cases. It shows that majority of voltage magnitude errors in zer-noise case lies in range of $[10^{-4}, 10^{-5}]$ pu. This range gets higher values of deviation and as the distribution curve moves to the right as the noise. As expected, exponential increase in noise level has resulted in a consistent, exponential boost of voltage magnitude deviation.

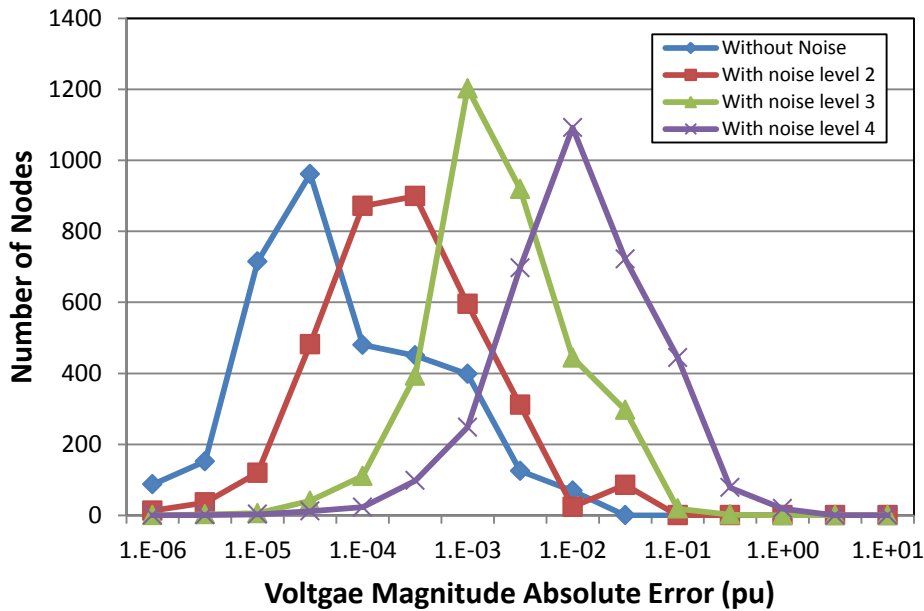

Fig. 17 – Histogram of the absolute errors of voltage magnitudes for different levels of noise

According to Fig. 18 which illustrates the histogram of the absolute errors of voltage magnitudes of the entire network for different cases, majority of the deviations of voltage phase are less than one degree in the cases with no noise and level-1noise. The consistent increase in deviation due to increase in level of noise is observed. The exponential increase of both noise level and deviation are also remarkable.

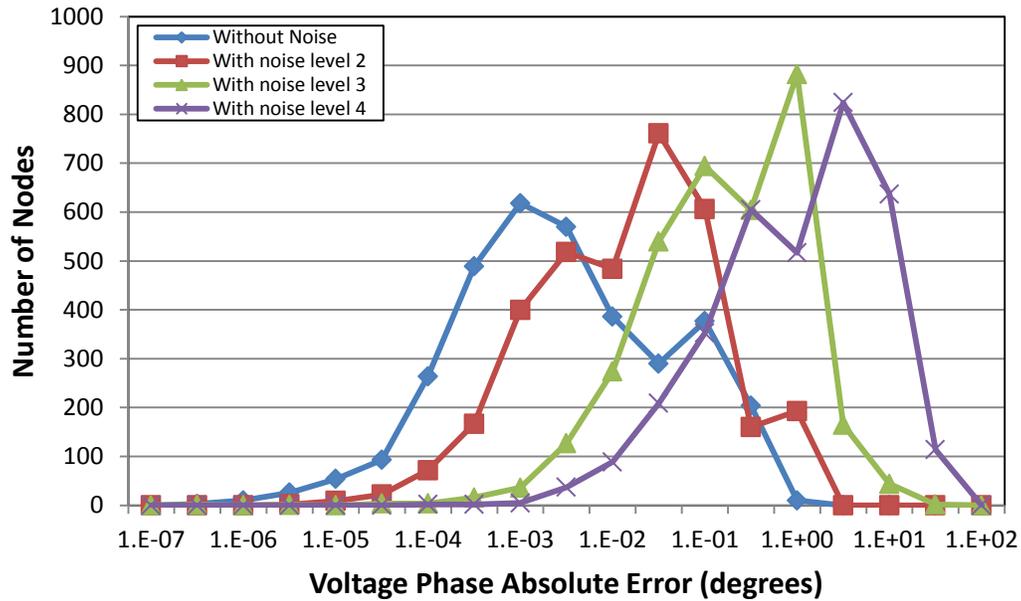

Fig. 18 – Histogram of the absolute errors of voltage phase for different levels of noise

# Conclusion

In this report, SDP method is employed to solve both PSSE and DSSE problems. Due to larger number of state variables in SDP-based SE compared to Newton's method, the SDP-SE method must be equipped with some algorithms to improve observability, noise cancellation, and bad data detection/identification. An algorithm has been developed in this report to generate additional measurements using the available measurements to improve observability. In order to eliminate the errors caused by noise corruption in huge networks, a network separation algorithm has been developed to divide the power network to smaller sub-networks including $\mu$PMUs. A redundancy test method has been also developed for bad data detection/identification.

The SDP-based PSSE is tested against single phase IEEE Standard 399, 30-bus, and 37-bus test systems. The results demonstrate the capability of the method to refine the correct solution in presence of noise and bad data in measurements. The SDP-based DSSE is applied to multiphase IEEE 13-bus, 37-bus, and 123-bus test feeders, which has proved the capability of the algorithm to solve the SE problem for such networks. The bad effects of noise on SE solutions are mitigated in IEEE 37-bus and 123-bus test feeders by employing decoupled SDP-based. The capability of the network separation algorithm is investigated on EPRI Circuit 5 test system which has 2998 bus and 3437 nodes. Statistical results of SE problem in presence of different levels of noise demonstrate a meaningful relation between noise level and solution errors.